\documentclass{aa}  

\usepackage{graphicx}

\begin{document}

 
\authorrunning{L. \v C. Popovi\'c,  E. G. Mediavilla,
P. Jovanovi\'c, J. A. Mu\~noz}
 
\titlerunning{The influence of microlensing on the Fe K$_\alpha$ line}  

\title{The influence of microlensing on the shape of the AGN Fe K$_\alpha$ line}

\author{L.\v C. Popovi\'c\inst{1,}\inst{2}, E. G. Mediavilla\inst{3}, P. 
Jovanovi\'c\inst{1,}\inst{2} \and J. A. Mu\~noz\inst{3}}


\institute{Astronomical Observatory, Volgina 7, 11160 Belgrade 74, Serbia
\and
 Isaac Newton Institute of Chile,
 Yugoslavia Branch \and Instituto 
de Astrof\'\i 
sica de Canarias E-382005, La Laguna,
Tenerife, Spain}

\date{Received 12 July 2002; accepted 14 November 2002}

\abstract{

We study the influence of gravitational microlensing on the AGN Fe K$_\alpha$
line confirming that unexpected enhancements recently detected in the iron line
of some AGNs can be produced by this effect. We use a ray tracing method to
study the influence of microlensing in the emission coming from a compact
accretion disc considering both geometries, Schwarzschild and Kerr.

Thanks to the small dimensions of the region producing the AGN Fe K$\alpha$
line, the Einstein Ring Radii associated to even very small compact objects
have size comparable to the accretion disc hence producing noticeable 
changes
in the line profiles. Asymmetrical enhancements contributing differently to the
peaks or to the core of the line are produced by a microlens, off-centered
with
respect to the accretion disc.

In the standard configuration of microlensing by a compact object in an
intervening galaxy, we found that the effects on the iron line are two orders
of magnitude larger than those expected in the optical or UV emission
lines.
In particular, microlensing can satisfactorily explain the excess in the iron
line emission found very recently in two gravitational lens systems, H 1413+117
and MG J0414+0534. 

Exploring other physical { scenario} for microlensing, we found that 
compact
objects (of the order of one Solar mass) which  belong to { the bulge or
the halo} of the host galaxy can also produce significant changes in the 
Fe K$_\alpha$ line profile of an AGN. However,  the optical depth estimated for
this type of microlensing is { very small, $\tau\sim 0.001$, even in a 
favorable case.}

}

\maketitle

\section{Introduction}

The influence of microlensing on the line profile of Active Galactic
Nuclei (AGNs)  has been
discussed by several authors.  On the basis of the standard model for AGNs
it
was accepted that the region generating the Broad Emission Lines (BELs)
was too large to be affected by microlensing by a Solar mass star
(Nemiroff 1988, Schneider \& Wambsganss 1990). However, recent studies
indicate that the BLR size is smaller than supposed in the standard model
(Wandel et al. 1999, Kaspi et al. 2000). According to this, Popovi\'c et
al (2001a)
considered the influence of microlensing on the spectral line profile
generated by a relativistic accretion disc in the Schwarzschild geometry,
finding that significant changes in the line profile can  be induced by
microlensing. Abajas et al. (2002) noted that this effect can be very
strong for high ionization lines arising from the inner region of the
accretion disc and identified a group of ten gravitational lens
systems in which microlensing of emission line region could be
observed. The scope of these studies has been the region where the
broad UV and optical lines are generated. However, microlensing detection
should be much more favorable in a compact region generating the X-Ray
radiation (Popovi\'c et al. 2001b).

The X-Rays of AGNs are generated in the innermost region of an accretion disc
around a central super-massive Black Hole (BH). An emission line from iron
K$\alpha$ (Fe K$\alpha$) has been observed at 6-7 KeV in the vast majority of
AGNs (see e.g. Nandra et al. 1997, Fabian et al. 2000). This line is probably
produced in the very compact region near the BH of an AGN (Iwashawa et al.
1999, Nandra et al. 1999, Fabian et al. 2000) and can bring essential
information about the plasma
conditions and the space geometry around the BH. Thus it seems clear that the
Fe K$\alpha$ line can be strongly affected by microlensing and recent
observations of two lens systems seem to support this idea (Oshima et
al. 2001,
Chartas et al. 2002).

According to this, the aim of this paper is to investigate the influence of
microlensing on the AGN Fe K$_\alpha$ line shape originated in a compact
accretion disc around non-rotating and rotating BH. In \S 2 we discuss the
construction of an accretion disc image from photons traveling in a rotating
(Kerr) and non-rotating (Schwarzschild) black hole space-time geometry,
including microlensing by a compact object. In \S 3 we analyze the consequences
and signatures of gravitational microlensing. In \S 4 the effects of
microlensing in the Fe K$_\alpha$ line are explored in two different physical
scenarios.

\section{Observed flux from a compact accretion disc affected by microlensing}

The effects of microlensing on a compact accretion disc have been 
previously
analyzed by Popovi\'c et al. (2001b) and Chartas et al. (2002). However,
these
authors do not strictly include the effects of ray bending caused by the black
hole that significantly distorts the image of the accretion disc.  To take
into account this effect, we will use the ray tracing method
(Bao
et al. 1994, Bromley et al. 1997, Fanton et al. 1997, \v
Cade{\v z} et al. 1998) considering only those photon trajectories that reach
the sky plane at a given observer's angle $\theta_{\rm obs}$.  We will adopt
the analytical approach  proposed by \v Cade{\v z} et al. (1998). If $X$ and
$Y$ are the impact parameters that describe the apparent position of each point
of the accretion disc image on the celestial sphere as seen by an observer at
infinity, the amplified brightness is given by

\begin{equation}
I_p=\varepsilon(r)g^4(X,Y)\delta(x-g(X,Y))A(X,Y) 
\end{equation}
where
$x=\nu_{\rm obs}/\nu_0$ ($\nu_0$ and $\nu_{\rm obs}$ are the transition
and observed frequencies, respectively); $g=\nu_{\rm em}/\nu_{\rm obs}$
($\nu_{\rm em}$ is the emitted frequency from the disc), and
$\varepsilon(r)$ is the emissivity in the disc,
$\varepsilon(r)=\varepsilon_0\cdot r^{-q}$. $A(X,Y)$ is the amplification
caused by microlensing. We will consider two different approximations to
estimate this quantity.

\subsection{Microlensing by an isolated compact object}

In this case the amplification is given by the relation (see e.g. Narayan
\& Bartelmann 1999):

\begin{equation}
A(X,Y)={u^2(X,Y)+2\over{u(X,Y) 
\sqrt{u^2(X,Y)+4}}},
\end{equation}
where  $u(X,Y)$  corresponds to the angular separation between
lens and source in Einstein Ring Radius (ERR) units and  is
obtained from
\begin{equation}
u(X,Y) ={\sqrt{(X-X_0)^2 +(Y-Y_0)^2}\over \eta_0},
\end{equation}
$X_0,Y_0$ are the coordinates of   the microlens with respect to the disc
center (given in gravitational radii), and
$\eta_0$ is the Einstein Ring Radius (ERR) expressed in gravitational
radii.

The total observed flux is given by

\begin{equation}
F(x)=\int_{\rm image} I_p(x)d\Omega
\end{equation}
where $d\Omega$ is the solid angle subtended by the disc in the
observer's sky and the integral extends over the whole (line) emitting
region.

\subsection{Microlensing by a straight fold caustic}

In most of cases we can not simply consider that microlensing is caused by an 
isolated compact object but we must take into account that the
micro-deflector
is located in an extended object (typically, the lens galaxy). In this case,
when the size of the ERR of the microlens is larger than the size of the
accretion disc, we can describe the microlensing in terms of the crossing
of the
disc by a straight fold caustic (Schneider et al. 1992). The
amplification at a point close to the caustic is given by (Chang \& Refsdal
1984),
\begin{equation}
A(X,Y)=A_0+{K\over\sqrt{\kappa(\xi-\xi_c)}}\cdot
H(\kappa(\xi-\xi_c)),
\end{equation}
where $A_0$ is the amplification outside  the caustic,
$K=A_0\beta\sqrt{\eta_0}$ is the
caustic amplification factor, where $\beta$ is constant of order of unity
(e.g. Witt et al. 1993). $\xi$ is the distance perpendicular to the
caustic in gravitational radii units  and $\xi_c$ is the minimum distance 
from the disc center to the caustic. Thus,
\begin{equation}
\xi_c={\sqrt{X_c^2+Y_c^2}},
\end{equation}

\begin{equation}
{\rm tg}\alpha={Y_c\over X_c},
\end{equation}
 and 
\begin{equation}
\xi=\xi_c+{{(X-X_c){\rm
tg}\phi+Y_c-Y}\over{\sqrt{{\rm tg}^2\phi+1}}},
\end{equation}
 where
$\phi=\alpha+{\pi/2}$. $ H(\kappa(\xi-\xi_c))$ is the Heaviside function,
$H(\kappa(\xi-\xi_c))=1$, for $\kappa(\xi-\xi_c)>0$,  otherwise it is
0. $\kappa$ is
$\pm 1$, it
depends on the direction of caustic motion; 
if the direction of the caustic motion is from approaching side
of the disc $\kappa=-1$, otherwise it 
is +1. Also, in the special case of caustic crossing perpendicular to the
rotating
axis $\kappa=+1$ for direction of caustic motion from -Y to +Y, otherwise 
it is -1.

\begin{figure}
\includegraphics[width=8.8cm]{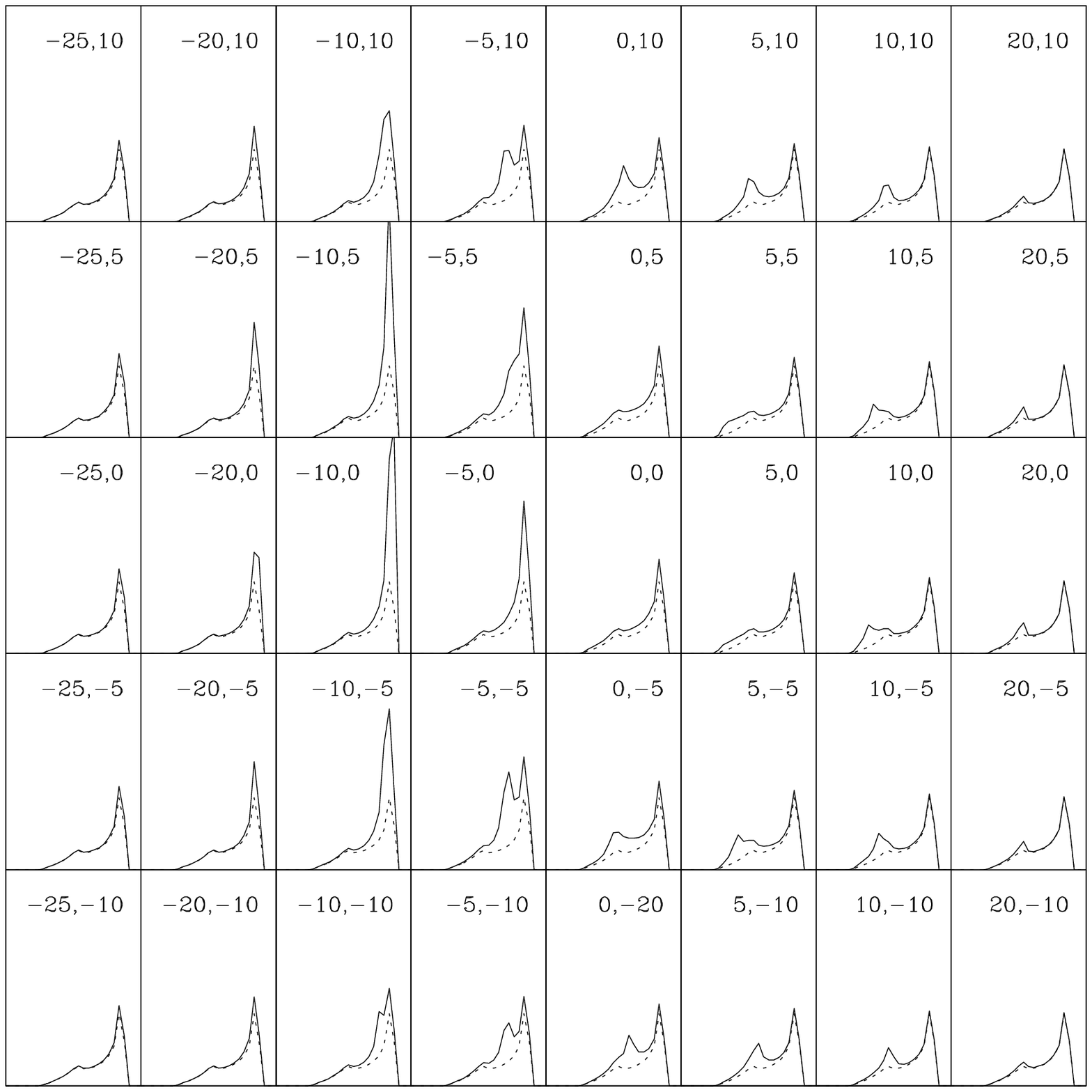}
\caption{The Fe K$_\alpha$ line profiles for different positions of  the
microlens.
 The calculations were performed for a
disc in the Schwarzschild metric with parameters: R$_{in}=6\ R_g$,
R$_{out}=20\
R_g$, $i=35^\circ$, and q=2.5. The ERR of the microlensing object is 10
R$_g$.
The relative intensity is in the range from 0 to 3, and $\nu/\nu_0$ is in
the range from 0.4 to 1.2. The numbers at the top of the figures are
the positions of the microlens (X$_0$,Y$_0$) with respect to the disc
center in gravitational radii.}
\end{figure}
 
\begin{figure}
\includegraphics[width=8.8cm]{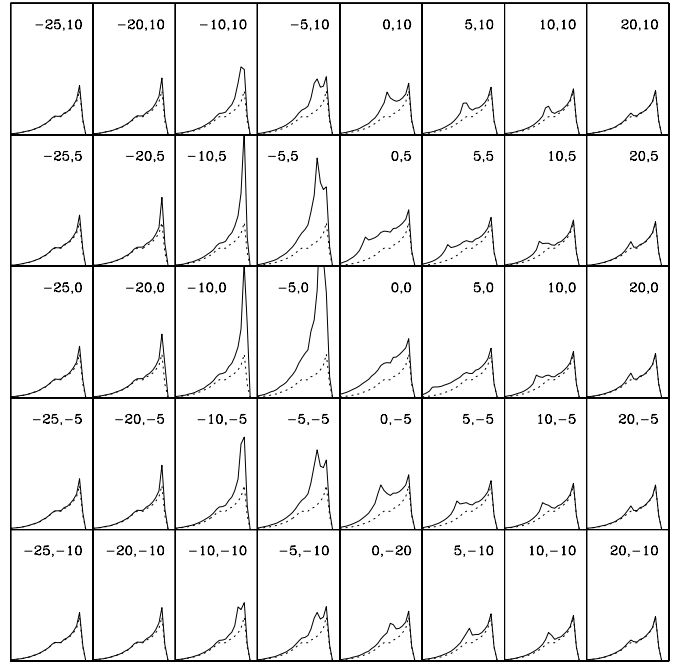}
\caption{The same as in Fig. 1, but for a rotating Kerr BH with $a=0.998$.
} \end{figure}

\section{Results}

To study the influence of MLE on the Fe K$\alpha$ line we adopt for the disc parameters the averaged
values given by Nandra et al.
(1997) from the study of 18 Seyfert 1 galaxies: $i=30^\circ$ and $q=2.5$ . For the inner radius we take
$R_{in}$=$R_{ms}$, where $R_{ms}$ is the radius of the marginal stability
orbit, that corresponds to $R_{ms}=6R_g$ in the Schwarzschild metric and to
$R_{ms}=1.23R_g$ in the case of the Kerr metric with angular momentum
$a=0.998$. Considering that for the adopted emissivity ($q=2.5$), the
emission is concentrated in the innermost part of the disc, we adopt for the
outer radius  $R_{out}=20R_g$.

\subsection{Microlensing by an isolated compact object}

We have computed the amplified line profile for different locations
with respect  to the center of the disc of a microlens of projected
Einstein
radius $\eta_0= 10\ R_{g}$. In Figs. 1 and 2 we show the results for
the Schwarzschild and Kerr metrics,  respectively. The distortions of the
line shape remind of the ones obtained in the case of the optical disc 
(Popovi\'c et al. 2001a) but the effects of microlensing are stronger in
the case of the X-Ray disc even when  we consider two order of magnitude
smaller microlenses. Notice also that the  inclination that we have
adopted for the disc is relatively low  ($i=30^\circ$) and that for
higher values of this parameter the effects will be considerably
increased (see Fig. 3).

Several outstanding changes of the line shape with the location of the
microlens,
and consequently with the transit of a microlens across the disc, can be
inferred
from Fig. 1 and 2. In the first place the number of peaks, their
relative
separation and the peak  velocity could change (this also affects to the
velocity
centroid). In the second place, the transit of a microlens would induce an
asymmetrical  enhancement of the line profile. For both metrics the
amplification
has a maximum  for negative values of $X_0$ that correspond to the
approaching part
of the rotating  disc. This amplification affects mainly  the blue part
of the
line, although  the asymmetrical enhancement induced by microlensing is
more
blue-ward in the Schwarzschild than in the Kerr metric.

\begin{figure}
\includegraphics[width=8.8cm]{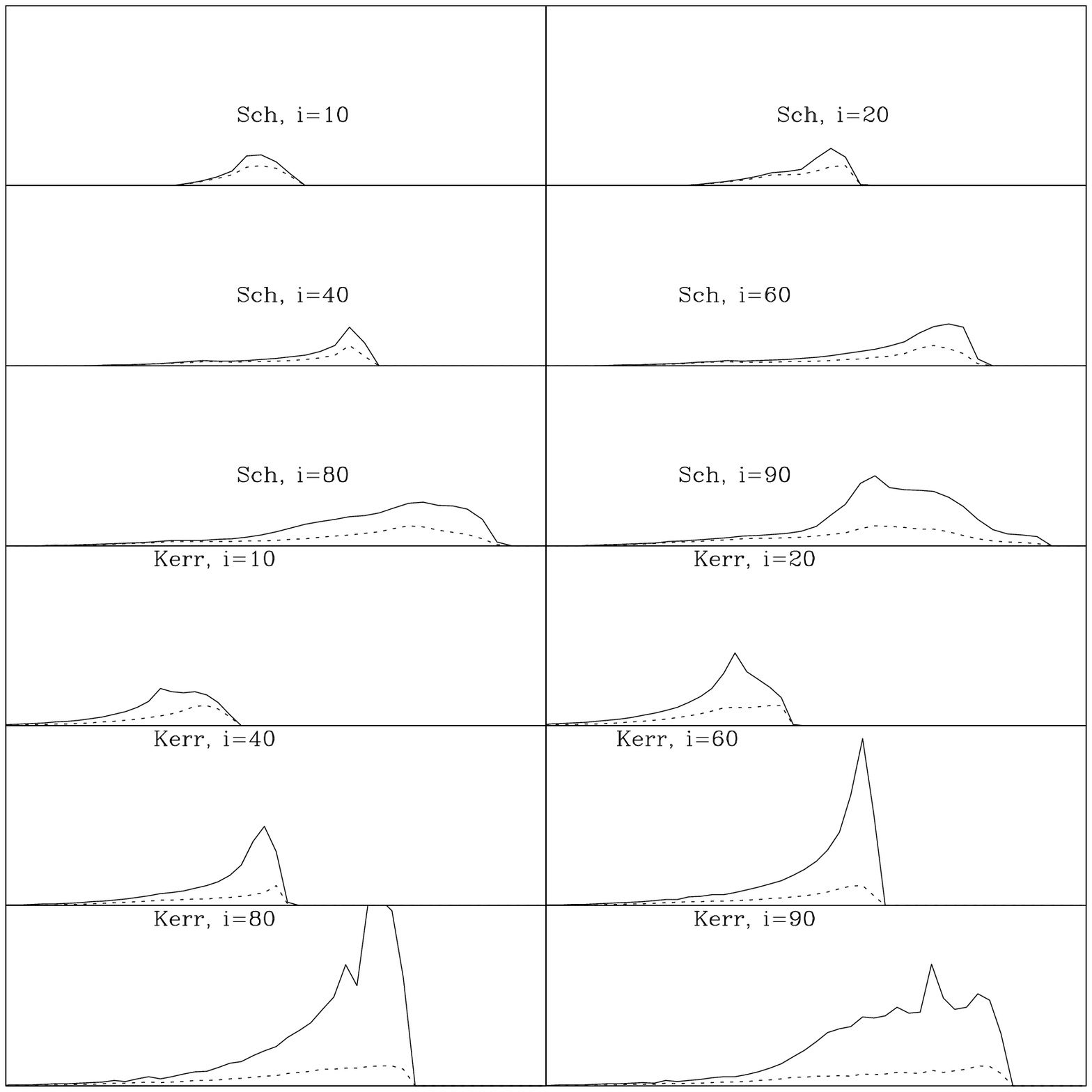}
\caption{ Influence of inclination in the deformation of the Fe K line shape (dashed line) induced by MLE (solid line): REE=10 R$_g$, $X_0=-5,\ Y_0=5$, for the
different
inclination ($i$) of the  disc in
Schwarzschild (Sch) and Kerr (Kerr)
metrics. The relative intensity is in the range from 0 to 10 (the maximum 
of unperturbed Fe K$_\alpha$ line is normalized to 1), and
$\nu/\nu_0$ is from 0.4 to 1.8}
\end{figure}

\subsection{Microlensing by a straight fold caustic}

In this case we have used the following parameters in Eq.(5): $A_0=1, \
\beta=1$ and ERR=50$R_g$.  We have studied the caustic crossing from different
orientations finding  results similar to the ones obtained in the previous
section for the  isolated microlens. In Figs. 4 (Schwarzschild) and 5 
(Kerr)
we show the  changes induced in the line profile by a caustic crossing 
perpendicular (first two sets of lines) and along (the third and fourth sets of
lines) the rotation axis starting from both sides of the disc, i.e. for
$\kappa=\pm 1$ respectively. In Figs. 4 and 5 we also present the
corresponding variations of line flux (below the sets of lines) for the whole
line (solid line) and for the blue ($\nu/\nu_0$ is 0.4-0.9, dashed line,
--- --- ---), central
(0.9-1.0, -- -- --) and red (1.0-1.2, - - -) parts of the line. As we can see
the caustic crossing can produce significant amplification of the line flux.
 As in the case of the
isolated microlens, the amplification is higher in the Kerr metric.  { 
This is in
agreement with previous investigation by Jaroszy\'nski et al. (1992) for
continuum radiation of the disc in the Schwarzschild and Kerr metrics.} 
The line
enhancement is also stronger in the blue part of the line, with this
asymmetrical trend more marked in the Schwarzschild case. 

\begin{figure*}
\includegraphics[width=18cm]{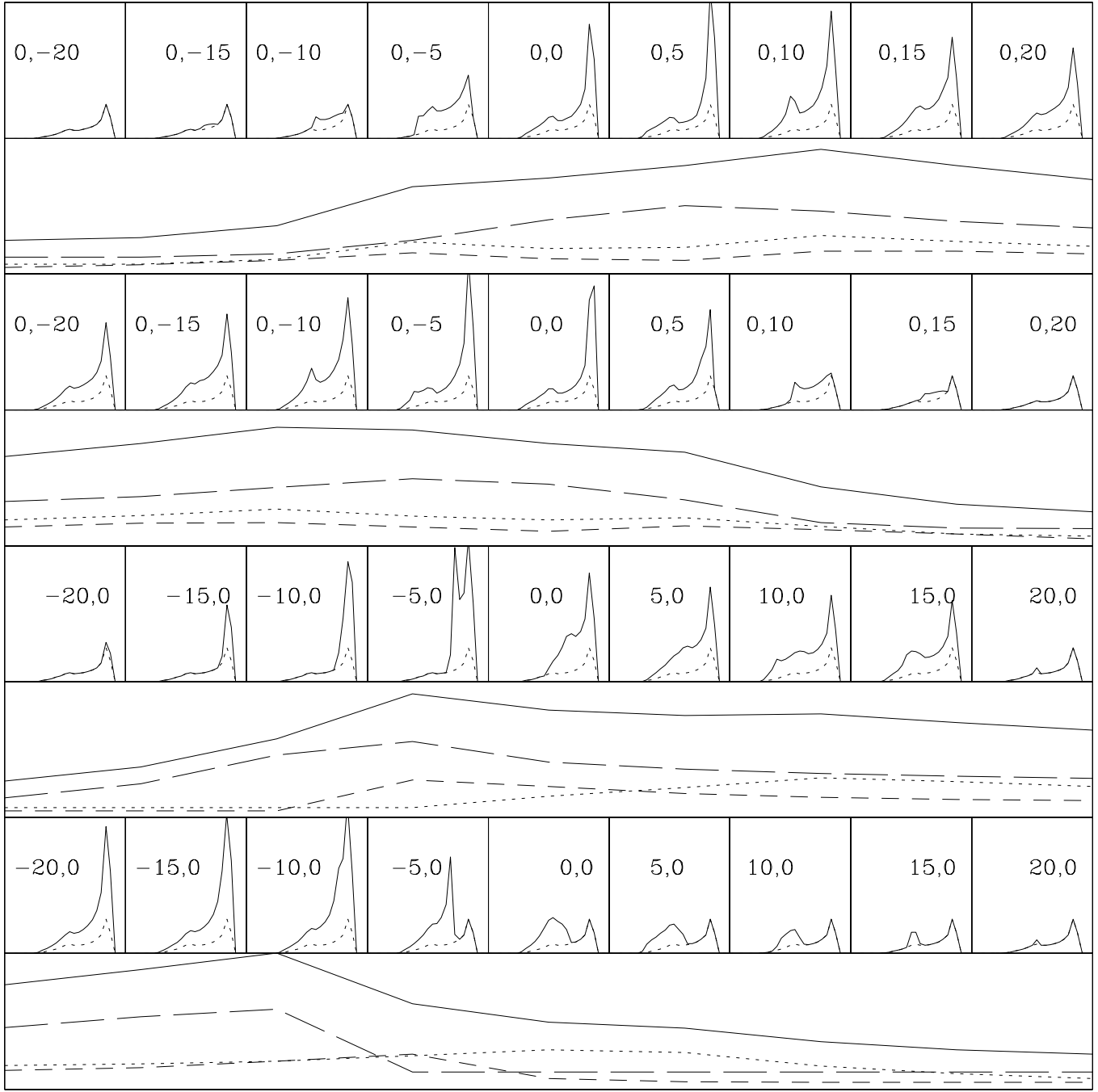}
\caption{ The caustic crossing the disc with the same parameters as in Fig.
1. The first and second sequences of lines present the crossing of caustic
perpendicular to the rotating axis for $\kappa=\pm 1$, respectively.
 The third and fourth sequences of lines present the caustic crossing along
the rotation axis for $\kappa=\mp 1$, respectively. Below the sequences
the
corresponding line flux variation is  presented:
solid line presents the whole line flux variation, dashed lines present
the variation of the red (0.4-0.9, - - -), central (0.9-1.0, -- -- --) and
blue (1.0-1.2, --- --- ---) parts of the line.  The relative intensity
ranges from 0 to 4 and $\nu/\nu_0$ ranges from 0.4 to 1.2.
The parameters of the disc and caustic are given in the text}
\end{figure*}

\begin{figure*}
\includegraphics[width=18cm]{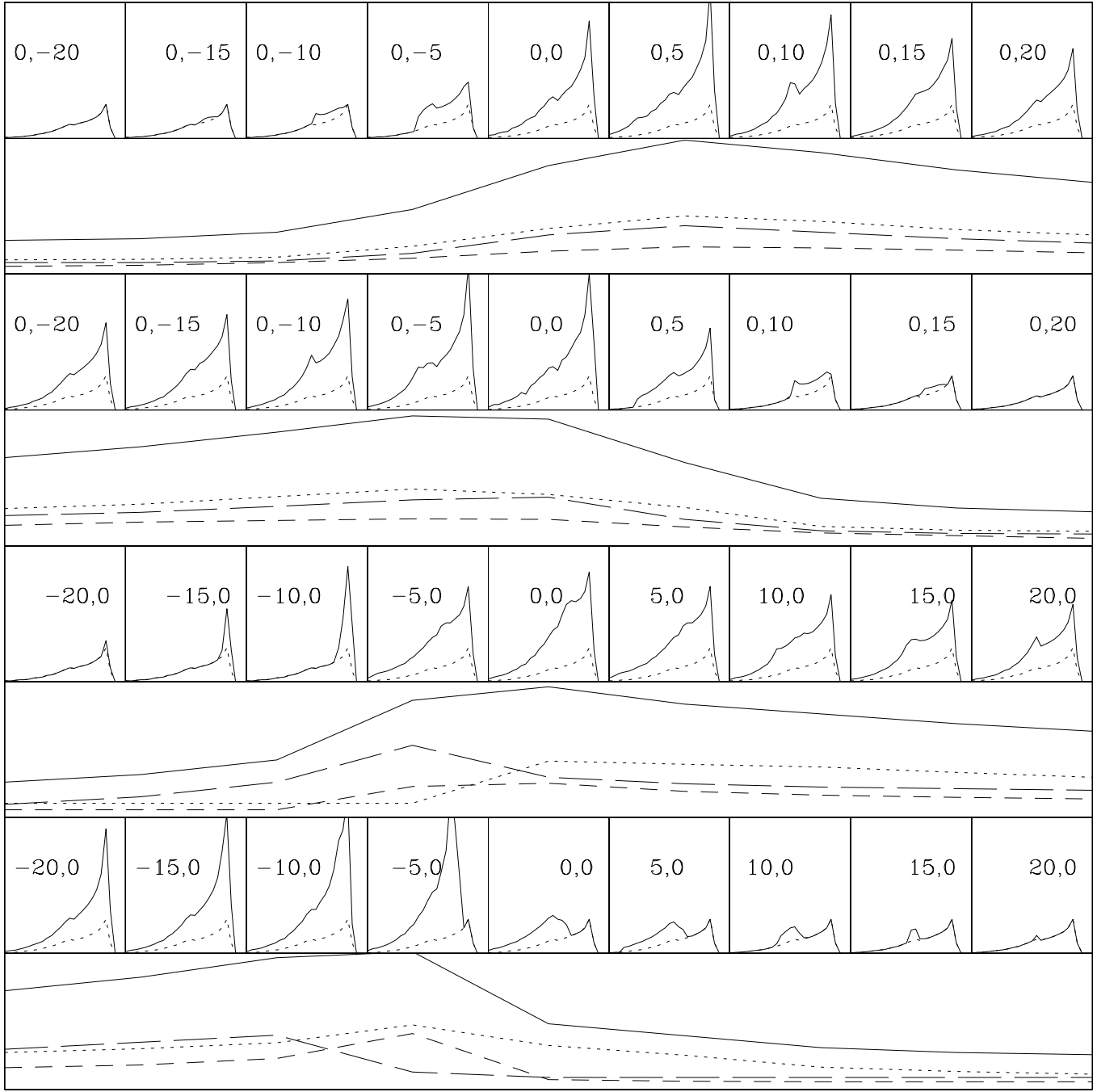}
\caption{ The
same as in Fig. 4 but for Kerr metric.}
\end{figure*}

\section{Physical scenarios for the microlensing of the AGN Fe K$_\alpha$
line}

In this section, we are going to explore two different physical scenarios
in which microlensing could induce detectable changes in the line profile
of the Fe K$\alpha$ line.

\subsection{Microlensing by a star-sized object in an intervening galaxy}

In the standard configuration, microlensing affects  one of the images of a
lensed QSO and is produced by a star-sized object in the lens galaxy. An event
of this type affecting  the Fe K$\alpha$ line has been reported in the
quadruple imaged QSO J0414+0534 by Chartas et al. (2002). These authors
observed a sudden increase in the iron line equivalent width from $\sim 190$ eV
to 900 eV only on image B of J0414+0534 proposing that it has been caused by a
microlensing event. We can try to reproduce this enhancement with our models
under the assumption of amplification by a straight fold caustic crossing.
However the problem is highly unconstrained because both sets of variables, the
one related to the microlensing (relative amplification, $\beta$, orientation
of the caustic with respect to the rotation axis, direction of the caustic
crossing, microlens mass) and the one related to the relativistic disc model 
(outer radius, emissivity, metric) should be considered to fit a unique number.
Thus, we can do only an exploration of scenarios compatible with the result. In
first place we can fix the disc parameters adopted until now to study which
values of the parameter $\beta$ can reproduce the observed amplification in
both metrics, Schwarzschild and Kerr. We have computed the maximum
amplification for a caustic crossing along the X axis. The resulting
amplifications, summarized in Table 1, indicate that there is a range of
$\beta$ values and microlens masses that can give rise to the observed
amplification. For instance, for the value $\beta=0.2$ (Chartas et al 2002), a
microlens of a solar mass can produce the observed enhancement. For $\beta=1$,
a low mass lens ($\sim 0.001\ M_\odot$) can also produce the required
amplification.


\begin{table}
\caption{The maximal amplification of the line flux for different
caustic   parameter
$\beta$ and masses of deflector ($M_{ml}$) for the case of MG
J0414+0534 and H 1413+117. The calculation was performed for
disc with the same characteristic as in the previous case, but for
$R_{out}=100\ R_g$. The calculations were performed for: a) Schwarzschild
and b) Kerr metric}

\vbox{\tabskip=0pt \offinterlineskip
\def\podvuci{\noalign{\hrule}}
\def\razmak{\noalign{\vskip.1truecm}}
\halign to \hsize {\strut#& \vrule#\tabskip=0pt  plus 10pt minus5pt &
\hfil#\hfil& \vrule#&
\hfil#\hfil&\vrule#&
\hfil#\hfil&\vrule#&
\hfil#& \vrule#
\tabskip=0pt\cr\podvuci
\razmak
\noalign{a) Schwarzschild metric}
\razmak
\podvuci
&&  M$_{ml}$ ($M_\odot$)  && $A_{\beta=0.2}$ &&
$A_{\beta=0.5}$ && $A_{\beta=1}$&\cr
\podvuci
&&   0.001                          &&  1.44 && 2.10&& 3.21& \cr
&&  0.01                          &&  1.78 && 2.95&& 4.91& \cr
&&   0.1                         &&  2.38 && 4.46&& 7.92&\cr
&&   1                          &&  3.47 && 7.16&& 13.33&\cr
\podvuci
\razmak
\noalign{b) Kerr metric}
\razmak
\podvuci
&&  M$_{ml}$ ($M_\odot$)  && $A_{\beta=0.2}$ &&
$A_{\beta=0.5}$ && $A_{\beta=1}$&\cr
\podvuci

&&   0.001                          &&  1.53 && 2.33&& 3.95& \cr
&&  0.01                          &&  1.94 && 3.36&& 5.73& \cr
&&   0.1                         &&  2.67 && 5.19&& 9.37&\cr
&&   1                          &&  3.98 && 8.45&& 15.91&\cr
\podvuci}}
\end{table}


In second place we can fix the relative microlensing amplification to
these values and try to fit the outer radii for several values of the
emissivity in both metrics, Schwarzschild and Kerr. The results are
included in Table 2. As it can be seen the results are relatively
independent of the outer radii; the emissivity being the more relevant
parameter.

\begin{table}
\caption{The maximal amplification of the line flux for different
disc   parameters
$R_{out}$ and q for the case of MG J0414+0534 and H 1413+117. The
calculation was
performed for caustic parameters: $\beta=0.2$, $M_{ml}=1M_\odot$,
$A_0=1$.}

\vbox{\tabskip=0pt \offinterlineskip
\def\podvuci{\noalign{\hrule}}
\def\razmak{\noalign{\vskip.1truecm}}
\halign to \hsize {\strut#& \vrule#\tabskip=0pt  plus 10pt minus5pt &
\hfil#\hfil& \vrule#&
\hfil#\hfil&\vrule#&
\hfil#\hfil&\vrule#&
\hfil#\hfil&\vrule#&
\hfil#& \vrule#
\tabskip=0pt\cr\podvuci
\razmak
\noalign{Schwarzschild metric}
\razmak
\podvuci
&&${q/R_{out}}$&&  20   &&  50 && 100&& 500 & \cr
\podvuci
&& 0 &&  4.3  &&  3.0 && 2.4 && 1.6& \cr
&& -1 && 4.4  && 3.1 &&2.6 && 1.7&\cr
&& -2 && 4.8  && 3.6 && 3.0 && 2.8&\cr
&& -3 && 5.2  && 4.8 && 4.3 && 4.1&\cr
&& -5 && 7.4 && 6.8 && 6.6 && 6.5 &\cr
\podvuci
\razmak
\noalign{Kerr metric}
\razmak
\podvuci
&& ${q/R_{out}}$&& 20 && 50 && 100 && 500 &\cr
\podvuci
&& 0 && 4.4 && 3.0  && 2.5 && 1.6 & \cr
&& -1 && 4.5 && 3.2 && 2.8 && 1.7 & \cr
&& -2 && 5.6 && 3.8  && 3.4 && 2.9 &\cr
&& -3 && 6.4 && 5.9 && 4.9 && 4.8 &\cr
&& -5 && 9.2  && 8.7 && 7.7 && 6.8 &\cr
\podvuci}}
\end{table}

Oshima et al. (2001) have also reported the
presence of a strong (EW$\sim 960 eV$) emission Fe K$\alpha$ line in the
integrated spectra of the the lensed QSO H 1413+117. Oshima et
al. (2001) interpret this results as
produced by
iron K$_\alpha$ emission arising from X-Ray re-processing in the broad
absorption line region flow.
Alternatively, we can assume that the individual spectra of the
components (not available) are different, and that the excess emission in
the iron line arises only from  one of the components, like in the case
of  J0414+0534. The red-shifts for the
source (2.558; Kneib et al. 1998) and the lens (0.9) in H 1413+117 
are so similar to the ones in J0414+0534 that we can use Tables
1 and 2 also for H 1413+117. The observed enhancement of one
order of magnitude (Oshima et al. 2001) in this object could be explained
by a 1$M_\odot$ if $\beta\sim 0.5$. 

Regardless of the true nature of the two events reported in J0414+0534 and H
1413+117, our analysis show that objects in a foreground galaxy with 
{ even relatively small masses} can bring strong changes in the line
flux. { This fact is indicating that changes in the Fe K$\alpha$ line 
flux will be higher than in the UV and optical lines}. Thus, the observation of 
the iron line in multi-imaged AGNs opens new possibilities to study the 
unresolved structure in QSOs and also the nature and distribution of compact
objects in lens galaxies.

\subsection{Microlensing by massive stars in the halo of an AGN}

The size of the ERR projected on the source, $\eta_0$ increases with the
distance between the source and the microlens. For this reason, appreciable
amplifications of the optical and UV BELs (i.e. $\eta_0$ comparable or greater
than the dimensions of the accretion disc) induced by a star-sized object are
only possible if the microlens is far away from the source, in an intervening
galaxy (typically the lens galaxy). In the optical and UV case, appreciable
amplifications of the BELs from an object in the host galaxy of the AGN are
possible only by a very massive object (Popovi\'c et al. 2001a). However, due
to the comparatively tiny dimensions of the X-Ray line emission region,
microlensing of a star-sized object in the host galaxy becomes a possibility.

Since in the { scenario} we are considering, the distance, $D$, 
between the
microlens and the accretion disc is negligible with respect to the distance
between the observer and the microlens, the expression for $\eta'_0$ can be
approximated as 
\begin{equation}
\eta'_0\approx\sqrt{{4GM_{ml}\over{c^2}}\cdot 
D},
\end{equation}
where $\eta'_0=\eta_0\cdot R_g$, and from Eq. (5)  the mass of the microlens
can be estimated as

\begin{equation}
M_{ml}[M_\odot]\approx {\eta_0^2\over{D}}\cdot
{G\over{4c^2}}M^2_{BH}=1.19\cdot
10^{-14}\cdot{\eta_0^2\over{D}}\ M_{BH}^2,
\end{equation}
 where $\eta_0$ is given in gravitational radii,  $D$ is given in parsecs
and $M_{BH}$ is given in solar masses.
 Selecting distances between the accretion disc and the microlens
in the interval 1 to 3 kpc (Schade et al. 2000),
we obtain masses for the microlens which are
smaller than solar
in the case of BH with mass $\sim 10^7$, of the order of several solar
masses for BH
mass of  $\sim 10^8$, and  for a very massive black hole $\sim
10^9M_\odot$
the microlens should have significantly higher masses than solar. In Table
3 the estimation of microlens masses for a distance of 1kpc from the disc
for three values of BH mass is given.

\begin{table}
\caption{The masses of the ML (in Solar masses) for different masses of
BH, the assumed distance
between deflector
and the accretion disc is 1kpc.}

\vbox{\tabskip=0pt \offinterlineskip
\def\podvuci{\noalign{\hrule}}
\def\razmak{\noalign{\vskip.1truecm}}
\halign to \hsize {\strut#& \vrule#\tabskip=0pt  plus 10pt minus5pt &
\hfil#\hfil& \vrule#&
\hfil#\hfil&\vrule#&
\hfil#\hfil&\vrule#&
\hfil#& \vrule#
\tabskip=0pt\cr\podvuci
&& ERR in $R_g$  && $M_{BH}=10^7$  && $M_{BH}=10^8$
 && $M_{BH}=10^9$ &\cr
\podvuci
&&   1                          &&  0.001 && 0.1&& 10 & \cr
&&   5                          &&  0.03 && 2.97 && 279& \cr
&&   10                          &&  0.12 && 12&& 1200&\cr
&&   20                          &&  0.48&& 48&& 4800&\cr
&&   50                         &&  2.97 && 297&& 29700 &\cr
\podvuci}}
               \end{table}

As one can see from the Table 3, a low-mass deflector from the bulge
of an AGN can produce significant changes in the line profile.

On the other side, the variation in the line profile and flux are very rapid in
the Fe K$\alpha$ line (see e.g. Vaughan \& Edelson 2001). To estimate the time
of microlens transition over the disc, we can use an approximative relation
$$\Delta t\approx {D_{disc}\over V}$$ where $D_{disc}$ is dimension of the
disc, and $V$ is velocity of microlens. As one can see from the Figs. 1 -- 
5,
the part of disc where the influence is strong and can be noticeable is of the
order of a few gravitational radii (or $10^{-6}-10^{-4}$ pc, if we take that
the velocity of microlens is of the order of $\sim 10^2$ km/s, we can estimate
that the corresponding time of variation is of the order of  $10^5-10^7$
seconds.

{ The probability of seeing a MLE is usually expressed in
terms of the optical depth $\tau$. As far as the potential microldeflectors
have an Einstein radius similar or larger than the radius of the
disc, $\tau$ can be estimated as the fraction of the area in the source 
plane 
covered by the projected Einstein radii of the microlenses. Taking also into
account that the distance between the accretion disc and the microlenses 
are
negligible as compared with the distance between the observer and the
microlenses

\begin{equation}
\tau\sim \frac{4\pi G}{c^2}\int_0^R \rho(r) r dr,
\end{equation}
where $R$ is the radius of the bulge or halo.

To estimate the order of magnitude of $\tau$
we assume a constant mass density, thus

\begin{equation}
\tau\approx \frac{2\pi G}{c^2}\rho_o R^2.
\end{equation}

Czerny et al. (2001) estimated that the total mass of the AGN bulges can
reach values of $10^{12} M_{sun}$ and Schade et al. (2000) found typical 
values
for the radius of the AGN bulges in the range 1--10 Kpc. Both results yield
to a maximum optical depth from the bulge $\tau_b\sim10^{-4}$.
Using also favorable numbers for the galactic halo 
($\rho\sim0.01 M_{\sun} pc^{-3}$, 
$R\sim 150 Kpc$) the optical depth could reach values of $\tau_h\sim 10^{-4}$. 
Adding both contributions a maximum optical depth $\tau\sim 0.001$ would be
expected in a favorable situation.
A detailed computation of $\tau$, in addition to include accurate density 
profiles, should
include a cut in the lower limit of integration ($r_{min}$) to exclude from
the integral the microdeflectors with Einstein radii smaller than the radius 
of the disc (although the approximation $r_{min}\sim 0$ used does not 
change 
significantly the order of magnitudes estimated for $\tau$ especially in the 
case of the halo).}

In spite of the relatively low probability of microlensing by a deflector in
the halo/bulge, we are going to explore the possibilities of detection in a
practical case.  We will study the case of NGC 3516.  This galaxy was 
continuously  monitored during 5 days by ASCA, and it was noted that the line
flux does vary on a short time-scale (Nandra et al. 1999). We consider the
Schwarzschild metric and a microlens with ERR of 10$R_g$ crossing along the
disc (with $Y_0$=0).

\begin{figure}
\includegraphics[width=8.8cm]{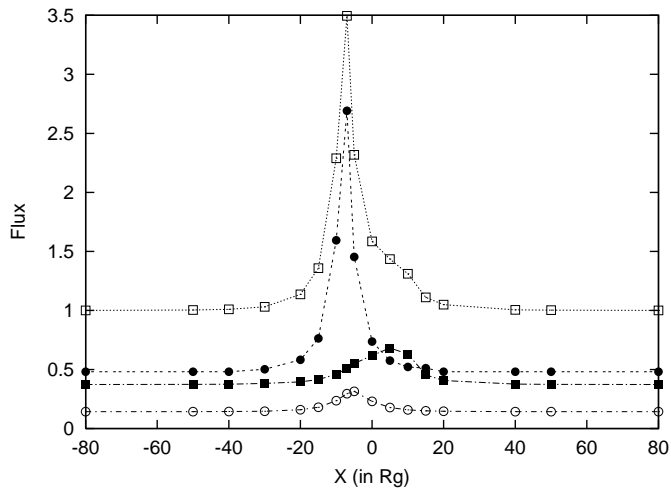} 
\caption{ The normalized flux variation in the central part (6.0 --
6.4 KeV, open circles),
blue wing
(6.4-7.2, full circles), red part (2.0-6.0, full squares) and total
line flux (open squares) as a function of position of the
microlens. Calculation was performed for   an accretion disc in
Schwarzschild
metric with parameters: $i=35^\circ$, $R_{inn}=60\ R_g$, $R_{out}=80\
R_g$, and q=8 (Nandra et al. 1999).} \end{figure}

The consequent variation of the flux of the Fe K$\alpha$ line is presented in
Fig. 6. The variation is higher in the blue wing, and even for the case of 
no
such compact disc considered above (80R$_g$), only a small (several R$_g$)
region is responsible for the flux line variation, because the emission is very
strongly concentrated in the inner disc. If we take that the BH mass is around
$4\cdot 10^7$ (Padovani et al. 1990), then we can estimate that the mass of a
microlens with ERR=10R$_g$ corresponds of 0.02M$_\odot$ at the distance of
1kpc. As one can see from the Fig. such a small mass object can amplify 
the
line emission by  a factor 3 in a relatively short lapse of time (microlens
crossing only several R$_g$). Moreover, the obtained results are in a good
qualitative  agreement with observations given by Nandra et al. (1999) where 
the line core seems to follow the continuum (very small changes in Fig. 
6),
while the blue wing is unrelated and shows a greater amplitude ($\sim$ 2).
 Moreover, the red wing presents very small variations, as in Nandra et
al. (1999) observations.

\section{Conclusions}

In order to discuss the amplification and the shape distortion of the
Fe K$\alpha$
line, we have developed a ray-tracing model to study the influence of
microlensing in a compact relativistic accretion disc.
 We consider both, the Schwarzschild and Kerr metrics. The main
conclusions of our work are:

1)  Microlenses of very small projected Einstein radii ($\sim$ 10
R$_g$) can
give rise to significant changes in the iron line profiles. The effects are two or three order of magnitude greater than the ones
inferred for the UV and optical lines (Popovi\'c et
al. 2001a). Off-centered
microlenses would induce strong asymmetries in the observed line
profiles.

2)  The effects of microlensing show differences in the Kerr and
Schwarzschild metrics,  the amplitude of the magnification being greater
in the Kerr metric. The transit of a microlens along the rotation axis of
the accretion disc would induce a strong amplification of the blue peak
in the Schwarzschild metric when the microlens was centered in the
approaching part. In the Kerr metric the amplification will be greater
but will not affect so preferentially  the blue part of the line. This
difference could be interesting to probe the rotation of an accretion
disc.

3)  Even objects of very small masses could produce observable
microlensing in
the iron K$_\alpha$ line of multiple imaged QSOs. We obtain that
microlenses of 1 solar
mass can explain the measured Fe K$\alpha$ line excess in the J0414+0534
and H
1413+117 lens systems.

4)  We have also found that stellar mass microlenses in the { halo/bulge}
of the host galaxy could produce significant changes in the iron lines of an
AGN. However, the optical depth is { low, $\tau\sim 0.001$,} even in
favorable cases.

\begin{acknowledgements}
This work is a part of the project P6/88 ``Relativistic and Theoretical
Astrophysics'' supported by the IAC and ``
Astrophysical Spectroscopy of Extragalactic Objects'' supported by the
Ministry of Science, Technologies and Development
of Serbia.
\end{acknowledgements}


\begin{thebibliography}{}

\bibitem[]{} Abajas, C., Mediavilla, E.G., Mu\~noz, J.A., Popovi\'c, L. \v C.,
Oscoz A., 2002,  ApJ, 576, 640.

\bibitem[]{} Bao G., Hadrava P., Ostgaard E., 1994, ApJ 435, 55

\bibitem[]{} Bromley B.C., Chen K., Miller W.A., 1997, ApJ 475, 57.


\bibitem[]{} \v Cade\v z A., Fanton C., Calivani M., 1998, New Astronomy
3, 647.

\bibitem[]{} Czerny, B., Nikolajuk, M., Piasecki, M.,   Kuraszkiewicz, J.,
2001 MNRAS, 325, 865.

\bibitem[]{} Chang, K.; Refsdal, S., 1984, A{\&}A 132,  168

\bibitem[]{} Chartas G., Agol E., Eracleous M., Garmire G., Bautz M.W.,
Morgan N.D., 2002,  ApJ, 568, 509.


\bibitem[]{} Fabian A.C., Iwashawa K.,  Reynolds C.S., Young A.J., 2000,
PASP 112, 1145

\bibitem[]{} Fanton C., Calivani M., Felice F., \v Cade\v z A., 1997, PASJ
49, 159.

\bibitem[]{} Iwashawa K., Fabian A.C., Young A.J., Inoue H., Matsumoto C.,
1999, MNRAS 306, L19.

\bibitem[]{} { Jaroszy\'nski M., Wambganss J., Pacz\'ynski B., 1992, 
ApJ 396, L65.}

\bibitem[]{} Kaspi S., Smith P.S., Netzer H., Maoz D., Jannuzi B.T.,
Giveon U., 2000 A\&{A} 533, 631.           

\bibitem[]{} Kneib, J.P., Allion, D., Pell\'o, R., 1998, A\&A, 339, L65.

\bibitem[]{} Nandra K., George I.M., Mushotzky R.F., Turner T.J. and
Yaqoob
T., 1997, Astrophys. J. 477, 602

\bibitem[]{} Nandra K., George I.M., Mushotzky R.F., Turner T.J. and
Yaqoob T., 1999, Astrophys. J. 523, L17

\bibitem[]{} Nemiroff, R.J., 1988, ApJ 335, 593.

\bibitem[]{} Narayan R., Bartelmann M., 1999, in Formation of Structure in
the Universe (Eds. A. Dekler, J.P. Ostriker) Cambridge University Press,
p. 360.

\bibitem[]{} Oshima T., Mitsuda K., Fujimoto R., Iyomoto N., Futamoto K.,
et al., 2001,   ApJ 563, L103.

 
\bibitem[]{} Padovani P, Burg R., Edelson R.A., 1990, ApJ 353, 438. 

\bibitem[]{}
Popovi{\'c}, L.\v C.,  Mediavilla, E.G., Mu\~noz J., 2001a, A{\&}A 378,
295.

\bibitem[]{}
Popovi{\'c}, L.\v C.,  Mediavilla, E.G., Mu\~noz J., Dimitrijevi\'c M.S.,
Jovanovi\'c P., 2001b, Serb. Aston. J. 164,
73 (Also, presented on GLITP Workshop on Gravitational Lens Monitoring,
4-6 June 2001, La Laguna, Tenerife, Spain).

\bibitem[]{} Schade D. J., Boyle B. J., Letawsky M., 2000, MNRAS 315, 498.

\bibitem[]{} Schneider, P. \& Wambsganss, J., 1990, A{\&}A 237, 42.

 \bibitem[]{}  Schneider, P., Ehlers, J., Falco, E. E., 1992,
Gravitational Lenses,  Springer-Verlag, Berlin
Heidelberg, New York.

\bibitem[]{} Vaughan S., Edelson R., 2001, ApJ 548, 694.

\bibitem[]{} Wandel, A., Peterson, B. M., Malkan, M. A.  1999, ApJ, 526,
579

\bibitem[]{} Witt H.J., Kayser R., Refsdal S., 1993, A{\&}A 268, 501.  

\end{thebibliography}
\end{document}